\documentclass[prb,twocolumn, superscriptaddress,aps]{revtex4-1}

\usepackage[english]{babel}
\usepackage[utf8]{inputenc}
\usepackage{amsthm,amsmath,amsfonts,amssymb,amscd,cases}
\usepackage{dsfont}
\usepackage[T1]{fontenc}
\usepackage{graphicx}   
\usepackage{epstopdf}
\usepackage[font=small,skip=8pt,justification=raggedright]{caption}

\usepackage{placeins} 
\usepackage{float} 

\newtheorem{theorem}{Theorem}

\newcommand{\R}{{\mathord{\mathbb R}}}
\newcommand{\one}{{\mathord{\mathds 1}}}

\begin{document}
\title{Quantum Systems at The Brink:\\
Properties of Atomic Bound States at The Ionization Threshold}
\author{Dirk Hundertmark}
\affiliation{Department of Mathematics, Institute for Analysis, Karlsruhe Institute of Technology, 76128 Karlsruhe, Germany, and Department of Mathematics, Altgeld Hall, University of Illinois at Urbana-Champaign, 1409 W. Green Street, Urbana, IL 61801, USA}
\author{Michal Jex}
\affiliation{Department of Physics, Faculty of Nuclear Sciences and Physical Engineering, Czech Technical University in Prague, B\v rehov\'a 7, 11519 Prague, Czech Republic and CEREMADE, Dauphine University, Place du Maréchal de Lattre de Tassigny, 75775 Paris Cedex 16, France}
\author{Markus Lange}
\affiliation{Department of Mathematics, University of British Columbia, 1984 Mathematics Road, Vancouver, BC, Canada V6T 1Z2}
\noaffiliation

\date{\today}

\begin{abstract}
We give a rigorous argument that long--range repulsion stabilizes quantum systems; ground states of such quantum systems exist even when the ground state energy is 
precisely at the ionization threshold. For atomic systems at the critical nuclear charge, our bounds show that the ground state falls off like $\exp(-c\sqrt{|x|})$ for large $|x|$. This is much slower than what the WKB method predicts for bound states with energies strictly below the ionization threshold. 
For helium type systems at critical nuclear charge, we show that our upper bounds are sharp. This rigorously confirms predictions by quantum chemists. 
\end{abstract}

\maketitle

Except for the famous Wigner-von Neumann potentials \cite{vonNeuWig29}, bound 
states of quantum systems are usually found below the energies of scattering states. 
The bound state energies and the scattering energies are separated 
by the ionization threshold. Above this threshold, the particles cease to be bound and move to infinity. Below the threshold, the binding energy, the
difference between the ionization threshold and the energy of the bound 
state, is positive and regular perturbation theory predicts that bound states 
are stable under small perturbations of the 
parameters describing the quantum system. 
That is, the energies might move a bit under small perturbations, but they do not suddenly disappear.

Imagine a parameter of the quantum system being tuned such that the energy of a bound 
state, e.g., the ground state energy, approaches the ionization 
threshold. At this critical value, the perturbation theory in the parameter breaks down and it is 
unclear what happens \emph{exactly at} this 
binding--unbinding transition: Does the bound state disappear, i.e., 
the quantum system spreads out more and more and dissolves or does the 
bound state still exist at the critical parameter and then suddenly disappear. 

This is the question we address here. 
We consider Schr\"odinger operators in atomic units of the form
\begin{equation}
\label{eq:intro}
	H = -\frac{1}{2m}\Delta-V_\lambda(x) + U(x) 
\end{equation}
where $ -\frac{1}{2m}\Delta$ is the kinetic energy, $U$ a non-zero repulsive part of the potential and $-V_\lambda$  an attractive part of the potential depending on a parameter $\lambda$. This operator describes one-particle models, however with slight modifications it can also describe interacting many-particle systems. 
The well-known WKB asymptotic, see also the work of Agmon\cite{Agm82}, shows that the eigenfunction 
corresponding to a discrete eigenvalue $E=E(\lambda)$ of the operator \eqref{eq:intro} falls off like 
\begin{equation*}
\exp\left(-\sqrt{2m\Delta E}|x|\right) \quad \text{for large }x 
\end{equation*}
where $\Delta E$ is the binding energy. This does not provide any useful information at critical coupling 
when $\Delta E=0$. 
Therefore, a new approach is needed. 

Our method presented here can be viewed as a higher order correction 
to the WKB method. It shows that the ground state at the ionization threshold exist and falls off like 
\begin{equation*} 
	\exp\left(-c|x|^{1/2} \right) \quad \text{for large }x 
\end{equation*}
for some explicit constant $c>0$ when the long--range repulsion is of Coulomb type. It easily generalizes to 
other types of long--range repulsion, see Theorem \ref{thm:one} below.
The underlying intuition is that if the binding energy approaches zero, the bound state can only disappear when it tunnels through the potential barrier to infinity.
\begin{figure}[h]
\includegraphics[width=.6\columnwidth]{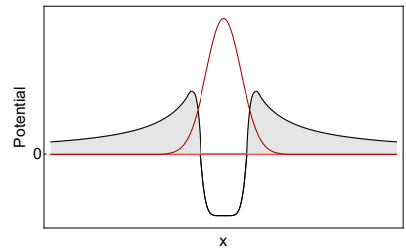}
\caption{Tunneling problem for the ground state at zero energy with sketched ground state.}
\centering
 \label{fig:tunneling}
\end{figure} 

If this tunneling probability is zero, the ground state cannot disappear, hence the quantum system stays 
bound at the critical coupling.  This is predicted by numerical calculations 
\cite{Hog95,DubIva98,KaiSer01,BusDraEstMoi14}.  Our approach makes this intuition precise, including upper 
bounds on the asymptotic behavior of the corresponding eigenfunctions at the ionization threshold.  
Moreover, we prove a strong dependence on the dimension in a one particle model: A 
repulsive part is only needed in dimensions $d\le 4$, with $d=4$ being critical,  while a repulsive part is 
not needed in dimensions $d\ge 5$.  This considerable generalizes previous results in three dimensions\cite{GriGar07}.

While we explain the main ideas in the one-particle model, a variety of 
physical systems can be handled. Particularly important are  
$N$ electron atoms with a nucleus of charge $Z$. 
For atomic systems, due to a classical result by Zhislin \cite{Zhi60}, ground states exist once $N<Z+1$. For $N > 2Z+1$, no such states exist\cite{Lie84}.
Hence, for fixed $N$ there is a critical charge $Z_c$ such that for $Z > Z_c$ bound states exist and for charges $Z < Z_c$ the quantum system has no bound state. 
Note that $Z_c$ does not have to be a whole number. 

For helium-like systems, a variational calculation of Bethe\cite{Bet29} shows that $Z_c<1$. Numerically, it is known\cite{BakFreDavHil} that $Z_c\sim 0.91$.
The existence and absence of the eigenstate for the simplest nontrivial example 
of helium-like system when $Z=Z_c$, was studied extensively by T.\ and M.\ Hoffmann-Ostenhof and Simon\cite{HofOstHofOstSim83}. They derived the existence of an eigenstate at a critical coupling $Z_c$ for a singlet state\cite{HofOstHofOstSim83}. 
Furthermore\cite{Hof84}, the triplet state case does not have a bound state at all and in this case, the critical coupling $Z_c=1$.

For general atoms, the existence of a ground state at critical coupling was studied in the  Born-Oppenheimer approximation\cite{BelFraLieSei14} and without it under the condition\cite{GriGar07} $Z_c\in(N-2,N-1)$. All these results establish the existence  
of an eigenstate, but the derived decay bounds are far from what is physically expected\cite{Hog98}.

Our approach relies mostly on energy estimates which, when combined 
with a geometrically inspired\cite{Uch69} lower bounds for the 
multiparticle potentials  of atomic systems, are also applicable to 
many-particle systems. \\
\linebreak
\textit{One-particle model} We explain our method using a one-particle model. Consider an eigenstate $\psi$ of the Hamiltonian $H$ from \eqref{eq:intro} associated with the eigenvalue $E$,  i.e.
\begin{equation*}
H\psi=E\psi\,.
\end{equation*}
For a suitable increasing function $F$, which diverges at infinity, we show that $e^{F}\psi$ is bounded in $L^2$-norm. This implies, that $\psi$ must asymptotically behave like $e^{-F}$, at least in $L^2$ sense. 

Denote by $\Sigma:=\lim_{R\rightarrow\infty}\Sigma_R$ the ionization threshold\cite{per60,CycFroKirSim87}, where
\begin{equation*}
\Sigma_R=\inf_{\|\psi\|=1,\mathrm{supp}\psi\subset\{|x|\ge R\}}\langle\psi,H\psi\rangle\,.
\end{equation*}
If $\chi_R$ is a function localizing smoothly in the region $\{|x|\ge R\}$, i.e., 
$0\le \chi_R\le 1$ and $\chi_R(x)=0$ for $|x|\le R$ while $\chi_R(x)=1$ for 
$|x|\ge 2R$, 
then using a variant\cite{Gri04} 
of the IMS localization formula\cite{Ism61,Mor79,Sig82}, we obtain
\small
\begin{align*}
\langle\chi_Re^F\psi,H\chi_Re^F\psi\rangle-\frac{1}{2m}\langle\psi,|\nabla(\chi_Re^F)|^2\psi\rangle =E\|\chi_Re^F\psi\|^2\,.
\end{align*}
\normalsize
The majority of terms in $|\nabla(\chi_Re^F)|^2$ are compactly supported. Denoting these terms by 
$G$ (the good part) and the remainder terms by $B$ (the bad part), we obtain
\begin{equation*}
	\left\langle\chi_Re^F\psi,\left(H-E-B\right)\chi_Re^F\psi\right\rangle\leq \|G\psi\|^2\leq K\,
\end{equation*}
where $K$ is a finite constant. The last step is to show that $H-E-B$ is positive. 
Here the repulsive part of the potential in $H$ is important since in the critical case $E\rightarrow\Sigma$. We summarize the conclusion of this procedure in the following theorem. 

Note that the existence of the ground state is a necessary assumption. However, using Tightness\cite{HunLee12}, the existence of an eigenstate for the critical case can be shown \cite{HunJexLan19-Helium}. 

\begin{theorem} \label{thm:one}
Each normalized eigenfunction $\psi$ corresponding to an eigenenergy $E \leq \Sigma$ of $H$ satisfies 
\begin{equation*}
	|\psi| \lesssim e^{-F - \frac{1}{2}\ln{\left(\Sigma-E+ U - \frac{|\nabla F|^2}{2m}\right)}}
\end{equation*}
for any function $F$ that satisfies for all $|x|\geq R>0$
\begin{equation*}
	\frac{|\nabla F|^2}{2m} < \Sigma-E+ U \,.
\end{equation*}
\end{theorem}
For the subcritical case, i.e., $E<\Sigma$, our result coincides with the result of Agmon \cite{Agm82} because 
\begin{equation*}
	\frac{|\nabla F|^2}{2m}=\Sigma-E-\epsilon < \Sigma-E+ U 
\end{equation*}
for $F(x)=\sqrt{2m(\Sigma-E-\epsilon)}|x|$. However, in contrast to the usual WKB asymptotic our bound provides detailed information on how well the quantum system is  localized at the critical coupling. Note that the logarithmic expression in the exponent corresponds to a polynomial correction of the asymptotic behavior. 
We note that the result in the theorem does not provide pointwise bounds for the function $\psi$. Nonetheless, it is possible to obtain pointwise information about $\psi$ under the additional assumption that $\psi$ is positive and continuous. This is done by using Harnack inequality \cite{AizSim82}. Recall that for a large class of Hamiltonians the ground state is always positive. The continuity follows from the fact that $\psi$ is in the domain of a differential operator.

\begin{figure}
\includegraphics[width=.95\columnwidth]{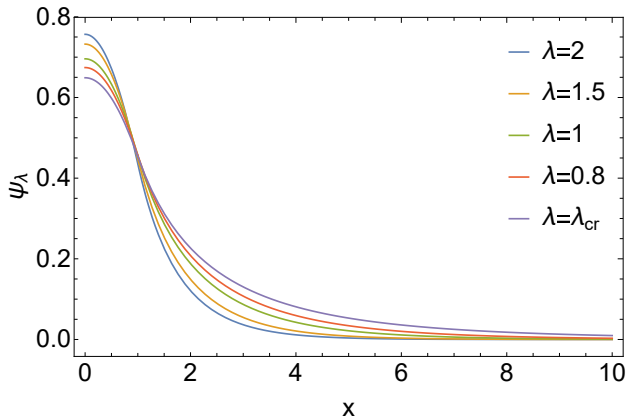}
\caption{Plot of normalized ground states for the Hamiltonian \eqref{eq:exam} with varying parameter $\lambda$.}
\centering
 \label{fig:ToyModel-longRangeCoulomb}
\end{figure}

As an illustrative example, consider the operator describing a quantum particle in a potential well with a long range Coulomb repulsion term present outside the well
\begin{equation}
\label{eq:exam}
H_\lambda=-\Delta-\lambda\,\one_{\{|x| \leq 1\}}+\frac{\one_{\{|x| > 1\}}}{|x|}\,.
\end{equation}
Here we chose $m=\frac 1 2$ for convenience. It can be easily shown that there exists a critical value $\lambda_{\mathrm{cr}}$ s.t. for $\lambda>\lambda_{\mathrm{cr}}$, the Hamiltonian $H_\lambda$ has bound states and for $\lambda< \lambda_{\mathrm{cr}}$ there are none. Furthermore, for this system we have $\Sigma=0$ and $\lambda_{\mathrm{cr}} \approx 0.634366
$. 
A plot of the ground state of $H_\lambda$ for a range of parameters $\lambda$ is given in Figure~\ref{fig:ToyModel-longRangeCoulomb}.

\begin{figure}
\includegraphics[width=.95\columnwidth]{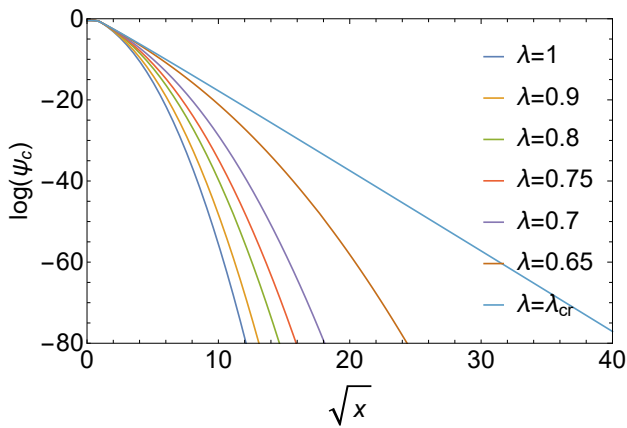}
\caption{Scaled plot of normalized ground states for the Hamiltonian \eqref{eq:exam} with varying parameter $\lambda$ for $x \in [0,1600]$. The convergence of the ground states for $\lambda \searrow \lambda_{\mathrm{cr}}$ is visible. Note that in this choice of scale the parabolic curves correspond to the ground state decays asymptotically as $\exp(-c|x|)$, as is predicted   by the WKB method, when the parameter 
$\lambda>\lambda_{\mathrm{cr}}$.  For $\lambda = \lambda_{\mathrm{cr}}$ the nearly straight line indicates that the ground state decays like $\exp(-\kappa \sqrt{|x|})$.}
\centering
 \label{fig:ToyModelDecay100}
\end{figure}

Using our theorem, we obtain for large $|x|$ and $\epsilon < 1$ the following upper bound for the ground state of $H_\lambda$ 
\begin{equation*}
\psi_{\lambda}(x)\lesssim e^{-\sqrt{|E(\lambda)|+\frac{1-\epsilon}{|x|}}|x|-\frac{1-\epsilon}{\sqrt{|E(\lambda)|}}\mathrm{arcsinh}\left(\sqrt{\frac{|E(\lambda)||x|}{1-\epsilon}}\right)}\,.
\end{equation*}
In the subcritical case, the first part of the exponent corresponds to exponential fall-off and the second one is the polynomial correction since 
$\mathrm{arcsinh}[y]=\ln(y+\sqrt{y^2+1})$. As $\lambda\searrow \lambda_{\mathrm{cr}}$, the ground state energy $E(\lambda)$ approaches $0$ and the behavior of the ground state changes to
\begin{equation*}
\psi_{\lambda_{\mathrm{cr}}}(x)\lesssim e^{-2\sqrt{(1-\epsilon)|x|}}\,.
\end{equation*}
See Figure~\ref{fig:ToyModelDecay100} for an illustration. 
A matching lower bound 
\begin{equation*}
N e^{-2\sqrt{(1+\epsilon)|x|}}\leq\psi_{\lambda_{\mathrm{cr}}}(x)\, 
\end{equation*}
can be obtained using a Comparison Lemma\cite{HofOst80-JPhysA}. Explicit calculations show that the eigenfunction has asymptotic behavior in the form
\begin{align*}
\psi_{\lambda_{\mathrm{cr}}}(x)\sim N\sqrt{\frac{\pi}{2}}\frac{e^{-2\sqrt{|x|}}}{2|x|^{3/4}}
\end{align*}
which is in perfect agreement with our result.

Our second example illustrates how the existence/non-existence of ground states at critical couplings depends on the dimension\cite{HunJexLan20-Potential}. For this we need the notion of a nonnegative operator, i.e., $H\ge 0$, if $\langle \psi, H\psi\rangle \ge 0$ for all functions $\psi$ in the domain of $H$. Equivalently, the spectrum of $H$, the set of allowed energies of the quantum system, is a subset of $[0\,,\infty)$. 
In addition, we say a potential $V$ is critical, if $H=-\Delta+V(x) \ge 0$ but, for any nonnegative function $W$ which is not identically zero, the perturbed operator 
$ H-W$ has a negative bound state. 

In the following, let $H = -\Delta+V(x)$ be nonnegative. 
\begin{theorem}[Non-existence of ground states at the ionization threshold]
$H$ does not have a ground state with energy $0$ if for some $R_0>0$ and all $|x|\ge R_0$ one has
\begin{equation*}
	V(x)\le \frac{d(4-d)}{4|x|^2} 
		+\frac{1}{|x|^2\ln x} 
		+ \frac{3}{4|x|^2(\ln x)^2} \,.
\end{equation*}
\end{theorem}

Complementary to that 

\begin{theorem}[Existence of ground states at the ionization threshold]
$H$ has a ground state with energy $0$ if $V$ is a critical potential and, for some $R_0>0$ and all $|x|\ge R_0$, one has 
\begin{equation*}
	V(x)\ge  \frac{d(4-d)}{4|x|^2} 
		+\frac{1+\epsilon}{|x|^2\ln x} 
		+ \frac{3+\epsilon}{4|x|^2(\ln x)^2} 
\end{equation*}
for $\epsilon>0$. 

\end{theorem}

Note that $\frac{d(4-d)}{4}$ is positive if $d\le 3$, zero if 
$d=4$, and negative if $d\ge 5$. 
Hence, in dimensions $d=2,3,4$ the potential $V$ has to have a \emph{positive tail} to support a ground state at the ionization threshold, whereas in dimensions $d\ge 5$ it can be negative.

To illustrate this, let $V$ be a simple potential well, i.e.,
\begin{equation*}
	V(x) :=-\one_{\{|x|\leq R\}}= \begin{cases} 
		-1\,, &\quad |x|\leq R\\
		0\,, &\quad |x|> R
	\end{cases}
\end{equation*}
 and consider the perturbed Hamiltonian 
$H_\lambda=-\Delta +\lambda V(x)$. It is well-known that $H_\lambda$ always has a bound state below the ionization threshold in dimension $d=1,2$\cite{ReeSim4}. Whereas in dimensions $d\ge 3$ the coupling $\lambda$ has to be big enough in order to have a ground state below zero\cite{ReeSim4}. If $\lambda$ decreases from large values, there will be a critical value $\lambda_c$ for which the ground state energy is zero. The question then is whether the ground state survives or disappears at $\lambda_c$. Our theorems above show that this depends crucially on the dimension. If $d\ge 5$ the ground state survives, whereas in dimensions $d<5$ it disappears. \\
\linebreak
\textit{Atoms} are described by the atomic Hamiltonian
\small
\begin{equation*}
H^{(N)}_{Z}=-\frac{\Delta_y}{2M}+\sum_{j=1}^N\left(-\frac{1}{2}\Delta_{x_j}-\frac{Z}{|x_j-y|}+\sum_{\substack{k=1\\ k< j}}^N\frac{1}{|x_j-x_k|}\right)
\end{equation*}
\normalsize
where $y$ corresponds to the position of the nucleus and $x_j$ are the positions of electrons. We are using atomic units, i.e., $\hbar=1$, $4\pi\epsilon_0=1$, $m_e=1$, and $e=1$. For convenience, we restrict to Born-Oppenheimer approximation\cite{BorOpp27}, i.e., a nucleus of infinite mass fixed at $y=0$,
\begin{equation}
\label{eq:hamiltonian}
H^{(N)}_{Z}=\sum_{j=1}^N\left(-\frac 1 2\Delta_{x_j}-\frac{Z}{|x_j|}+\sum_{\substack{k=1\\ k< j}}^N\frac{1}{|x_j-x_k|}\right)\,.
\end{equation}
However, our results also hold for the operator without this approximation\cite{HunJexLan19-Helium}. We denote the ground state energy of $H^{(N)}_{Z}$ by $E^{(N)}_{Z}$. 
In the first part, we saw that the asymptotic behavior in the critical case is controlled by the repulsive part of the potential. For atoms there are effectively two regions with different behavior. 
\linebreak

\noindent
\textit{Helium} For $N = 2$, 
we introduce for $\frac 1 2\leq \alpha\leq 1$ and $\delta > 0$ the region
\begin{align*}
A_{\delta,\alpha} = \{(x_1,x_2) \in \R^6\,:\, |x|_0 \geq \delta|x|_\infty^\alpha\}
\end{align*}
where $|x|_\infty := \max\{|x_1|, |x_2|\}$ is the distance  between the nucleus and the further away electron and $|x|_0 := \min\{|x_1|, |x_2|\}$ is the distance of the closer electron to the nucleus. We note that for the special case $\alpha=1$ we require $\delta<1$. 
Roughly speaking, in the region $A_{\delta,\alpha}$ we do not see the effective long range repulsion in $|x|_\infty$ and outside of  the region $A_{\delta,\alpha}$ we see only effective long range repulsion in $|x|_\infty$. This corresponds to the intuitive explanation that the electron closer to the nucleus shields it from the second electron. 
Inside and outside of the region we get the following lower bound for the potential from Eq.~\eqref{eq:hamiltonian} with $N=2$: 
\begin{align*}
&A_{\delta,\alpha}\!: U(|x_1|,|x_2|)
				\geq\left( \frac 1 2-Z\right) \frac{1}{|x|_\infty}-\frac{Z}{\delta|x|_\infty^\alpha}\,,\\
&A_{\delta,\alpha}^c\!: U(|x_1|,|x_2|)
		\geq - \frac{Z}{|x|_0} + \left(\frac{1}{1+ \delta|x|^{\alpha-1}_\infty}-Z\right)\frac{1}{|x|_\infty}\,. 
\end{align*}
Using these estimates we can bound the action of the operator \eqref{eq:hamiltonian} on a function $\psi_{A_{\delta,\alpha}}$ supported inside $A_{\delta,\alpha}$ and a function $\psi_{A^c_{\delta,\alpha}}$ supported outside $A_{\delta,\alpha}$. 
More precisely, inside the region $A_{\delta,\alpha}$ we can bound the kinetic energy of the electrons by zero and obtain a lower bound that depends only on $Z$ and $|x|_\infty$. Outside of $A_{\delta,\alpha}$ we have to deal with a potentially singular term since $|x|_0$ is no longer bounded away from zero. However, we  can bound this term with help of the corresponding kinetic energy term from below by $E_Z^{(1)}$. 
Plugging these estimates into the energy estimate, we can show that $H_Z^{(2)}-E_Z^{(1)}-B$ is 
positive provided that the function $F$  is chosen correctly in the two regions. 
 In particular, we can prove for constants  
$1>\epsilon_0>\frac 1 2>\epsilon_\infty> 0$ and $K_{0},K_{\infty}>0$ 
the following asymptotic behavior of an eigenstate at the threshold:
\begin{theorem}  A ground state $\psi$ of $H^{(2)}_{Z_c}$ falls off faster than
\footnotesize
\begin{align*}
	&	\exp\left(\!-\sqrt{2\left|E^{(1)}_{Z_c}\right|}|x|_0+K_0|x|_0^{\epsilon_0} -\sqrt{8(U - 1)} \,\sqrt{|x|_\infty}+K_\infty|x|_\infty^{\epsilon_\infty}\right) 
\end{align*}
\normalsize
 in the interior of the region $A_{\delta,\alpha}$ and faster than 
 \begin{align*}
	&	\exp(-\sqrt{8(U - 1)} \,\sqrt{|x|_\infty}+K_\infty|x|_\infty^{\epsilon_\infty})
\end{align*}
outside of $A_{\delta,\alpha}$.
\end{theorem}
It is important to note that the constant $\sqrt{8(U - 1)}$ holds only for $\frac 1 2<\alpha<1$. In the critical case $\alpha=1$, the constant becomes $\sqrt{8\frac{U}{1+\delta} - 8}$ which forces $\delta$ to be sufficiently small, namely, $\delta<U-1$. Moreover, the second 
critical case $\alpha=\frac 1 2$ has a constant $\sqrt{8\left(U-1-\frac{L}{\delta^2}\right)}$, where $L>0$. The main ingredients in the proof are again an application of IMS localization formula and a clever choice of cutoff functions to separate the coordinate space into manageable regions \cite{HunJexLan19-Helium}.

The lower bound for the Helium atom can be constructed in the same way as for one-particle model using Comparison Lemma. Unfortunately, we are only able to construct worse bounds compared to the one-particle case. In particular, we construct two lower bounds. The first one provides an exponential lower bound everywhere. The second one yields a subexponential lower bound outside of the regions $A_{\delta,\alpha}$ and $|x|>R>0$. The main difficulty in the construction of subsolutions for the operator \eqref{eq:hamiltonian} are singularities of the function $\frac{1}{|x_1-x_2|}$. These can be remedied by a polynomial correction term \cite{HofOst80-PhysLett} at the cost of a faster exponential fall-off constant.  For a positive ground state function of the operator Eq.~\eqref{eq:hamiltonian} such that 
$H^{(2)}_{Z_c}\psi = E^{(1)}_{Z_c} \psi$  and suitable constants  $n,R,N,C> 0$, we 
obtain for every $(x_1,x_2)\in\mathbb R^6$ that satisfies $|x|_\infty>R$
\begin{equation*}
	\psi\geq NM_n(|x_1-x_2|)\exp\left(-\sqrt{2\left|E^{(1)}_{Z_c}\right|}|x|_0-C|x|_\infty\right)
\end{equation*}
where the polynomial correction is given by 
\begin{equation*}
	M_n(\tau) := \begin{cases} 
		(\tau+n)^{n}\,, \quad &\tau\leq n\\
		(t(\tau)+n)^{n}\,, \quad &n<\tau<3n\\
		(3n)^{n}\,, \quad& 3n\leq \tau\\
	\end{cases}
\end{equation*}
for suitably chosen function $t(\tau)$. However, the issue with the singularities is only present within the region $A_{\delta,\alpha}$. Therefore, we can derive a much better lower bound outside of that region. 
\begin{theorem} 
For every $x\in A_{\delta,\alpha}^c$ s.t. $|x|_\infty>R$ the ground state $\psi$ of $H^{(2)}_{Z_c}$ satisfies
\small
\begin{equation*}
	\psi\geq N\chi\left(\frac{|x|_0}{\delta|x|_\infty^\alpha}\right)\exp\left(-\sqrt{2\left|E^{(1)}_{Z_c}\right|}|x|_0-c_1|x|_\infty^\frac{1}{2}-\eta|x|_\infty^\epsilon\right)
\end{equation*}
\normalsize
where $N,c_1,R,\alpha,\eta,\delta,\epsilon> 0$ and 
\begin{equation*}
	\chi(s) := \begin{cases} 
		1\,, &\quad 0\leq s\leq 1\\
		\cos\left(\frac\pi 2(s-1)\right)\,, &\quad 1<s<2\\
		0\,, &\quad 2\leq s\,.\\
	\end{cases}
\end{equation*}
Thus, inside of the region, in which intuitively one electron shields the nucleus from the other one, we obtain comparable lower and upper bounds. We note that for $\alpha\in(1/2,1)$ we have $c_1=\sqrt{8(U-1)}$.
\end{theorem}

\noindent
\textit{General atom} The procedure and the result for the upper bound can be generalized for an arbitrary atom with $N$ electrons. The main idea, i.e., construct two regions and estimate the action of the operator within, remains the same since there are still two regions with different asymptotic behavior. 
In the first one, all electrons are far away from the nucleus and the wavefunction falls off like an exponential function. 
In the second region, at least $N-K$ electrons remain relatively close to the nucleus while the other $K$ electrons can move far away. 
In this case, the $K$ outmost electrons are still bound but the wavefunction asymptotically behaves subexponentially. 
The physical mechanism for this confinement is not the attraction to the nucleus, which gets shielded by the inner electrons anyway,  but the  \emph{remaining long-range repulsion} which acts as an effective barrier that prevents the system to break up. Hence, at least a stretched exponential fall-off persists.

Let $E_Z^N$ be the ground state energy of an $N$-electron atom of nuclear charge $Z$. 
Unlike the 2 electron Helium type system, it can now happen that more than one electron can be removed without energy cost: 
we assume the existence of a critical coupling $Z_c<N-K$ such that  
\begin{equation*}
	E^{(N)}_{Z_c}=\ldots=E^{(N-K)}_{Z_c}<E^{(N-K-1)}_{Z_c}\,.
\end{equation*}
Thus the binding energy of $K$ electrons to the nucleus is equal to $0$, but the binding energy $\Delta E^{(N-K-1)}_{Z_c}$ of the $(K+1)$-th  electron is positive. Under the condition $N-2<Z_c<N-1$ it was proved that a ground state exists\cite{GriGar07}. While this seems to be a reasonable assumption\cite{Hog98,Ser99}, it is not rigorously known. We considerably strengthen the previously known results\cite{GriGar07, BelFraLieSei14} by establishing much better 
bounds on the fall-off of the ground state eigenfunction at critical charge.

For simplicity, we consider only the case where each electron can uniquely be identified by its distance to the nucleus, i.e. 
\begin{equation*}
|x|_1 < |x|_2 < \ldots < |x|_N\,, \quad \forall x\in\R^{3N}\,
\end{equation*}
where $|x|_k:\R^{3N}\rightarrow\R^+_0$ gives the $k$-th smallest value out of the distances $|x_j|$ of the electrons. This is not a real restriction. It is  equivalent to omitting a set of measure $0$ which can be treated rigorously \cite{HunJexLan19-NAtom}. Due to this simplification we can introduce unique coordinates $\tilde{x}_j$ such that for each fixed point $x\in\R^{3N}$ we have $|\tilde x_j|=|x|_j$. 
One can now easily write the regions and estimates needed for our method. A detailed exposition of the derivation is given in the supplementary material. The final estimate which is needed in the proof of the claim can be summarized as 
 \begin{equation*}
\begin{split}
&A_> := \{x \in \R^{3N} : |x|_{N-K}>\delta|x|_{N-K+1}\} :\\
&\quad U\geq U_{N-K-1}-\frac{Z}{\delta|x|_{N-K+1}}-\sum_{j=N-K+1}^{N}\frac{Z}{|x|_j}\,,\\
&A_< := \{x \in \R^{3N} : |x|_{N-K}<\delta|x|_{N-K+1}\}:\\
&\quad U\geq U_{N-K}+\sum_{j=N-K+1}^{N}\left(\frac{N-K}{1+\delta}-Z\right)\frac{1}{|x|_j}
\end{split}
\end{equation*}
where
\begin{equation*}
U_M:=\sum_{j=1}^M\left(-\frac{Z}{|x|_j}+\sum_{\substack{k=1\\ k< j}}^M\frac{1}{|\tilde x_j-\tilde x_k|}\right)\,.
\end{equation*}
Note that $U_M$ describes the interaction between a nucleus and ``inner'' $M$ electrons including interactions between them. The expressions are in spirit the same as in the case of Helium. In particular, for a function $\psi_>$ supported within the region $A_>$, the action of the operator is bounded from below by $\Delta E^{(N-K-1)}_{Z_c}$. 
Moreover, in the second region $A_<$ we obtain a positive lower bound for the action that depends on the distance of the outer coordinates to the nucleus.  To summarize:  \begin{theorem}  
There exists a small enough $\tilde{\delta}>0$ and suitable  constants  $C_j,K_j\in\R_+$ such that a ground state $\psi$ of $H^{(N)}_Z$ falls off like
 $$\exp\left(-\sum_{j=N+1-K}^NC_j\sqrt{|x|_j}\right)$$ 
 if $|x|_{N-K} < \tilde{\delta}|x|_{N-K+1}$ and like
\begin{align*}
	\exp\left(-\sum_{j=N+1-K}^NK_j|x|_j\right)
\end{align*}
otherwise. 
\end{theorem}

The proof\cite{HunJexLan19-NAtom} again reveals a distinct relation 
between the repulsive part of the potential and the asymptotic behavior 
of the eigenfunction similar to the one-particle model and also the two-particle (Helium) case.\\
\linebreak
In summary, we have shown, that existence and fall-off behavior of eigenfunctions at the critical coupling, for the class of operators \eqref{eq:intro}, 
depend on the asymptotic behavior of the potential. This was conjectured by quantum chemists.\cite{Hog98b}
We also provided explicit, dimension-dependent conditions for the potential under which a zero-energy ground states does, or does not, exist.

We demonstrated how to apply our method to atomic systems under the additional assumption that $N-K>Z_c$, where $Z_c$ is the critical charge of the nucleus, $N$ is the total number of electrons, and $K$ is the number of electrons leaving the atom as $Z$ decreases below $Z_c$. It does not require any symmetry restriction on the quantum particles. This means that our result is valid for any statistics imposed on the electrons in atoms. In a real physical system, electrons are fermions, which means that the ground state for more than 2 electrons can not be strictly positive anymore. This unfortunately implies that we can not use Comparison Lemma to obtain a lower bound.  It is necessary to find a different approach for systems with more than 2 electrons.

We only considered nonrelativistic quantum systems. For very large atoms, it is undoubtedly necessary to use, at least for the inner electrons, the corresponding relativistic equations to obtain the correct results. Our method relies mainly on the IMS localization formula. 
Thus using known results for pseudo-relativistic quantum systems\cite{BarHarHunSem19},  it should be possible to 
 to adapt our method to systems with pseudo-relativistic electrons.

Calculations suggest that similar results are valid within Hartree-Fock and Density Functional Theory (DFT). This is especially interesting due to the fact that these theories are inherently nonlinear. This would rigorously prove the asymptotic behavior predicted 
by various DFT-methods\cite{GB2015}. Another open problem is the  applicability of our method for the case of interacting systems of multiple atoms, i.e., molecules. 
The additional geometry, due to the relative positions of the multiple atoms, as well as a more complicated relation between energy and electron distribution make this a hard to tackle but also a very interesting problem. 
\vspace{-0.4cm}
\begin{acknowledgments}
\vspace{-0.3cm}
Dirk Hundertmark was funded by the Deutsche Forschungsgemeinschaft (DFG, German Research Foundation) – Project-ID 258734477 – SFB 1173. 
Michal Jex received financial support from the Ministry of Education,
Youth and Sport of the Czech Republic under the Grant No. RVO 14000. 
The work of Markus Lange was supported by NSERC of Canada. 
\end{acknowledgments}

\newpage  

\begin{center} 
{\LARGE Supplementary material} 
\end{center}

\noindent\textit{Lower bound estimate}
In order to prove our theorem in general setting we need to be able to estimate the potential
\begin{equation*}
U_N=\sum_{j=1}^N\left(-\frac{Z}{|x_j|}+\sum_{\substack{k=1\\ k< j}}^N\frac{1}{| x_j-x_k|}\right)\,
\end{equation*}
 from below in a suitable fashion. For this purpose we introduce the notion of inner and outer electrons as well as an ordering for the electrons. 
For simplicity we consider only the case where each electron can uniquely be identified by its distance to the nucleus, i.e. 
\begin{equation*}
|x|_1 < |x|_2 < \ldots < |x|_N\,, \quad \forall x\in\R^{3N}\,
\end{equation*}
where $|x|_k:\R^{3N}\rightarrow\R^+_0$ give the $k$-th smallest value out of the distances $|x_j|$ of the electrons.We introduce unique coordinates $\tilde{x}_j$ such that for each fixed point $x\in\R^{3N}$ we have $|\tilde x_j|=|x|_j$. 
We call the set\\
$\{\tilde x_k \,: \, k\in\{1,\ldots, N-K\}\}$\\
\hspace*{2cm} the inner coordinates and the set\\
$\{\tilde x_k \,: \, k\in\{N-K+1,\ldots, N\}\}$\\
\hspace*{2cm} the outer coordinates.
\\
Moreover the potential corresponding to the $M$ most inner electrons is denoted by
\begin{equation*}
U_M:=\sum_{j=1}^M\left(-\frac{Z}{|x|_j}+\sum_{\substack{k=1\\ k< j}}^M\frac{1}{|\tilde x_j-\tilde x_k|}\right)\,.
\end{equation*}
We begin estimating the potential similar to the two electron case
\begin{align*}
		&A_1\rightarrow |x|_1>\delta|x|_N:\\
		&\quad U\geq U_{N-K-1}-\left(1+\frac{K}{\delta}\right)\frac{Z}{|x|_N}\\
		&A_1^c\rightarrow |x|_1<\delta|x|_N:\\
				&\quad  U\geq  U_{N-1}+\sum_{j=2}^{N-1}\frac{1}{|\tilde x_j-\tilde x_N|}+\left(\frac{1}{1+\delta}-Z\right)\frac{1}{|x|_N}\,
\end{align*}
where we used $\frac{1}{|x_j-x_k|}>0$ and $-\frac{1}{|x|_j}>-\frac{1}{\delta|x|_N}$ within $A_1$ and $\frac{1}{|\tilde x_1-\tilde x_N|}>\frac{1}{|\tilde x_1|+|\tilde x_N|}>\frac{1}{(1+\delta)|\tilde x_N|}$ outside $A_1$. Provided that $N-K>1$ we split $A_1^c$ as follows
\begin{align*}
	&A_2\rightarrow |x|_1<\delta|x|_N\,\,\mathrm{and}\,\,|x|_2>\delta|x|_N:\\
		&\quad U\geq U_{N-K-1}-\left(1+\frac{K}{\delta}\right)\frac{Z}{|x|_N}\\
		&A_2^c\rightarrow |x|_2<\delta|x|_N:\\
				&\quad U\geq  U_{N-1}+\sum_{j=3}^{N-1}\frac{1}{|\tilde x_j-\tilde x_N|}+\left(\frac{2}{1+\delta}-Z\right)\frac{1}{|x|_N}\,.
\end{align*}
This process can be repeated for each inner coordinate, i.e. $N-K$ times. The last step yields
\begin{align*}
		&A_{N-K}\rightarrow |x|_{N-K-1}<\delta|x|_N\,\,\mathrm{and}\,\,|x|_{N-K}>\delta|x|_N:\\
		&\quad U\geq  U_{N-K-1}-\left(1+\frac{K}{\delta}\right)\frac{Z}{|x|_N}\\
		&A_{N-K}^c\rightarrow |x|_{N-K}<\delta|x|_N:\\
				&\quad U \geq  U_{N-1}+\hskip-5mm \sum_{j=N-K+1}^{N-1}\frac{1}{|\tilde x_j-\tilde x_N|}+\left(\frac{N-K}{1+\delta}-Z\right)\frac{1}{|x|_N}\,.
\end{align*} 

In the case that more than one electron leaves, i.e. $K > 1$, we can repeat the above procedure for all the remaining $K-1$ outer electrons.

The final estimate in the two resulting regions can be summarized as 
 \begin{equation*}
\begin{split}
&A_> := \{x \in \R^{3N} : |x|_{N-K}>\delta|x|_{N-K+1}\} :\\
&\quad U\geq U_{N-K-1}-\frac{Z}{\delta|x|_{N-K+1}}-\sum_{j=N-K+1}^{N}\frac{Z}{|x|_j}\,,\\
&A_< := \{x \in \R^{3N} : |x|_{N-K}<\delta|x|_{N-K+1}\}:\\
&\quad U\geq U_{N-K}+\sum_{j=N-K+1}^{N}\left(\frac{N-K}{1+\delta}-Z\right)\frac{1}{|x|_j}\,.
\end{split}
\end{equation*}

\begin{figure}[h]
\includegraphics[width=.95\columnwidth]{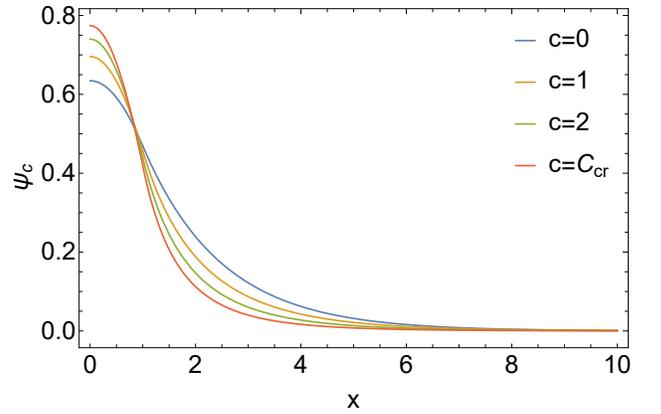}
\caption{Plot of the normalized ground state eigenfunction for the model \eqref{eq:exam_Supplementary} for several values of $c$.}
\centering
 \label{fig:ToyModel-longRangeCoulomb_supplementary}
\end{figure}

\noindent\textit{Numerical examples:} 
In the following we present additional examples of the asymptotic behavior for one-particle models. We consider again the operator describing a quantum particle in a potential well with a Coulomb repulsion term everywhere outside that well 
\begin{equation}
\label{eq:exam_Supplementary}
H_c=-\Delta-\one_{\{|x| \leq 1\}}(|x|)+\one_{\{ 1<|x| \}}(|x|) \frac{c}{|x|}\,.
\end{equation}
However we do not decrease the depth of the well but increase the repulsion term. Due to the increase of the repulsive term outside the eigenfunctions become more localized for $c \nearrow C_{\mathrm{cr}}$ where $C_{\mathrm{cr}} \approx 3.11693$ is the numerically calculated critical value, see Figure~\ref{fig:ToyModel-longRangeCoulomb_supplementary}.  

\begin{figure}
\includegraphics[width=.95\columnwidth]{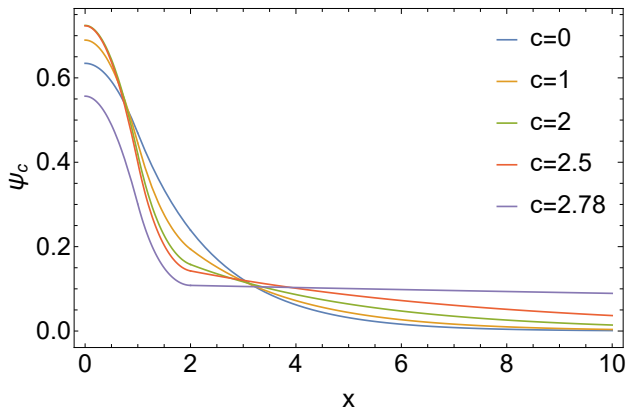}
\caption{Plot of the normalized ground state eigenfunction for the model \eqref{eq:NOlongRangeCoulomb} for several values of $c$. It illustrates, that as $c $ approaches $\widetilde{C}_{\mathrm{cr}}$ the wavefunction delocalizes. }
\centering
\label{fig:ToyModel-finitePotentialBarrier}
\end{figure}

In order to emphasis that it is crucial to have a long range repulsive term we now consider the following Hamiltonian
\begin{equation}
\label{eq:NOlongRangeCoulomb}
\widetilde{H}_{c}=-\Delta-\one_{\{|x| \leq 1\}}(|x|)+c\one_{\{1<|x|<2\}}(|x|)\,.
\end{equation}
The repulsive potential is present only in a finite region around the potential well. 
Note that the value $2$ is artificial and has no particular importance. 
If we start to increase the parameter $c$ up to the critical value $\widetilde{C}_{\mathrm{cr}} \approx 2.7938776$ we see that far away from the critical value the increase of $c$ leads to the localization of the wavefunction even by a short range potential. However for $c \geq 2.5$ the wavefunction starts to spread further and further and for $c = 2.78$ the fall-off of the function is hardly visible, see Figure~\ref{fig:ToyModel-finitePotentialBarrier}.

The presented plots highlight the physical intuition that the wavefunction has to tunnel through the repulsive barrier in order to leave the potential well and delocalize. 
However the long range Coulomb repulsion is too 'sticky' for the wavefunction to delocalize even at the critical value and hence we are able to prove fall-off behavior at the ionization threshold. \\

\noindent\textit{Depiction of fall-off regions for Helium Atom}
In order to give the reader a better understanding of the fall-off behavior at the threshold and in particular to illustrate the respective sizes of the different regions we plot the ground state behavior for Helium in the case $Z = Z_c$. 

\begin{figure}[H]
\includegraphics[width=.95\columnwidth]{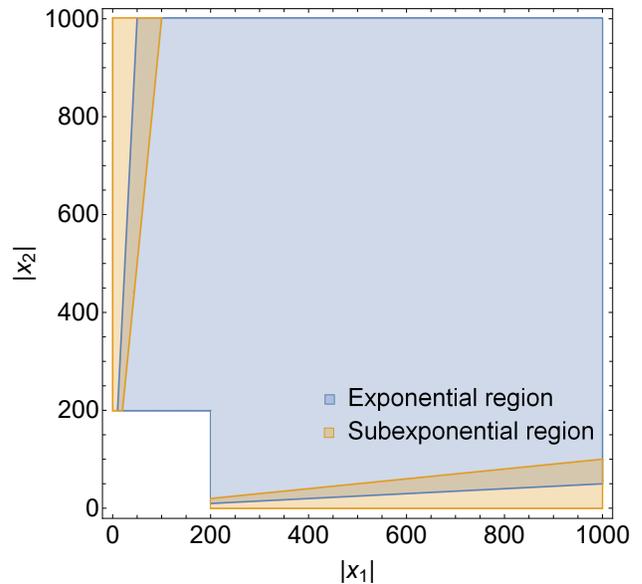}
\caption{Behavior of the ground state eigenfunction for Helium atom at the critical coupling for $\max\{|x_1|,|x_2|\}>200$ with the regions $\{(|x_1|,|x_2|) \,:\, \min\{|x_1|,|x_2|\} > 0.05 \max\{|x_1|,|x_2|\}\}$ (blue), $\{(|x_1|,|x_2|) \,:\, \min\{|x_1|,|x_2|\} < 0.1 \max\{|x_1|,|x_2|\}\}$ (amber).}
\label{fig:Helium_lin_true}
\end{figure}

\begin{figure}[h!]
\includegraphics[width=.95\columnwidth]{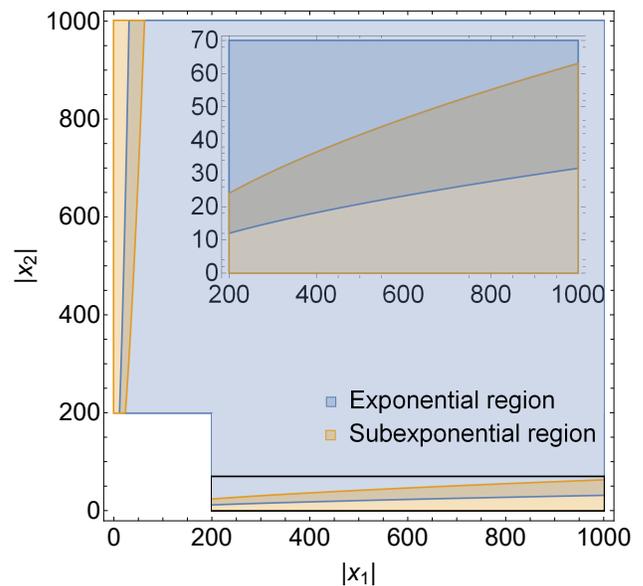}
\caption{Behavior of the ground state eigenfunction for Helium atom at the critical coupling for $\max\{|x_1|,|x_2|\}>200$ with the regions $\{(|x_1|,|x_2|) \,:\, \min\{|x_1|,|x_2|\} > 0.5 \max\{|x_1|,|x_2|\}^{0.6}\}$ (blue), $\{(|x_1|,|x_2|) \,:\, \min\{|x_1|,|x_2|\} <  \max\{|x_1|,|x_2|\}^{0.6}\}$ (amber). \newline The embedded graph is the zoomed in rectangular sector near the in $|x_1|$-axis.}
\centering
\label{fig:Helium_sub_true}
\end{figure}

Figures~\ref{fig:Helium_lin_true}~and~\ref{fig:Helium_sub_true} show that the exponential fall-off is almost everywhere. Only in the case that one of the electrons is close to the nucleus in comparison to the other one we obtain subexponential fall-off. This is due to the remaining long-range repulsion
which acts as an effective barrier that prevents the system to break up. Figures~\ref{fig:Helium_lin_true}~and~\ref{fig:Helium_sub_true} seem at the first glance similar. However the asymptotic behavior is significantly different. Volume of the amber cones grows as $\sim |x|_\infty^3$ for the case of Figure~\ref{fig:Helium_lin_true} and as $\sim |x|_\infty^{2.2}$ for the case of Figure~\ref{fig:Helium_sub_true}. 

\begin{figure*}
\begin{minipage}[t]{.45\textwidth}
\includegraphics[width=.95\columnwidth]{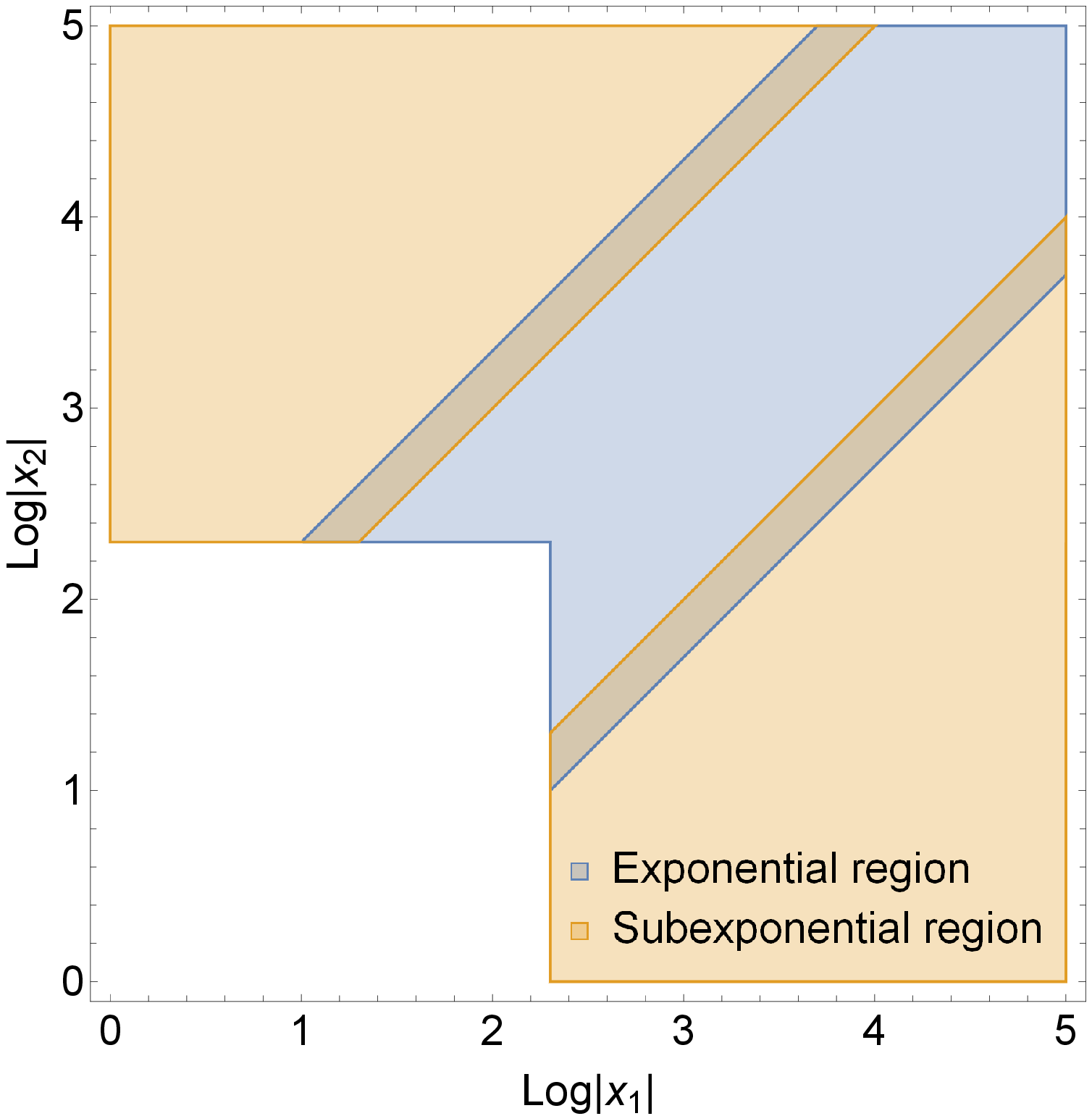}
\caption{Behavior of the ground state eigenfunction for Helium atom at the critical coupling for $\max\{|x_1|,|x_2|\}>200$ with the regions $\{(|x_1|,|x_2|) \,:\, \min\{|x_1|,|x_2|\} > 0.05 \max\{|x_1|,|x_2|\}\}$ (blue), $\{(|x_1|,|x_2|) \,:\, \min\{|x_1|,|x_2|\} < 0.1 \max\{|x_1|,|x_2|\}\}$ (amber).}
\centering
\label{fig:Helium_lin_log}
\end{minipage}\quad\quad\quad\quad
\begin{minipage}[t]{.45\textwidth}
\includegraphics[width=.95\columnwidth]{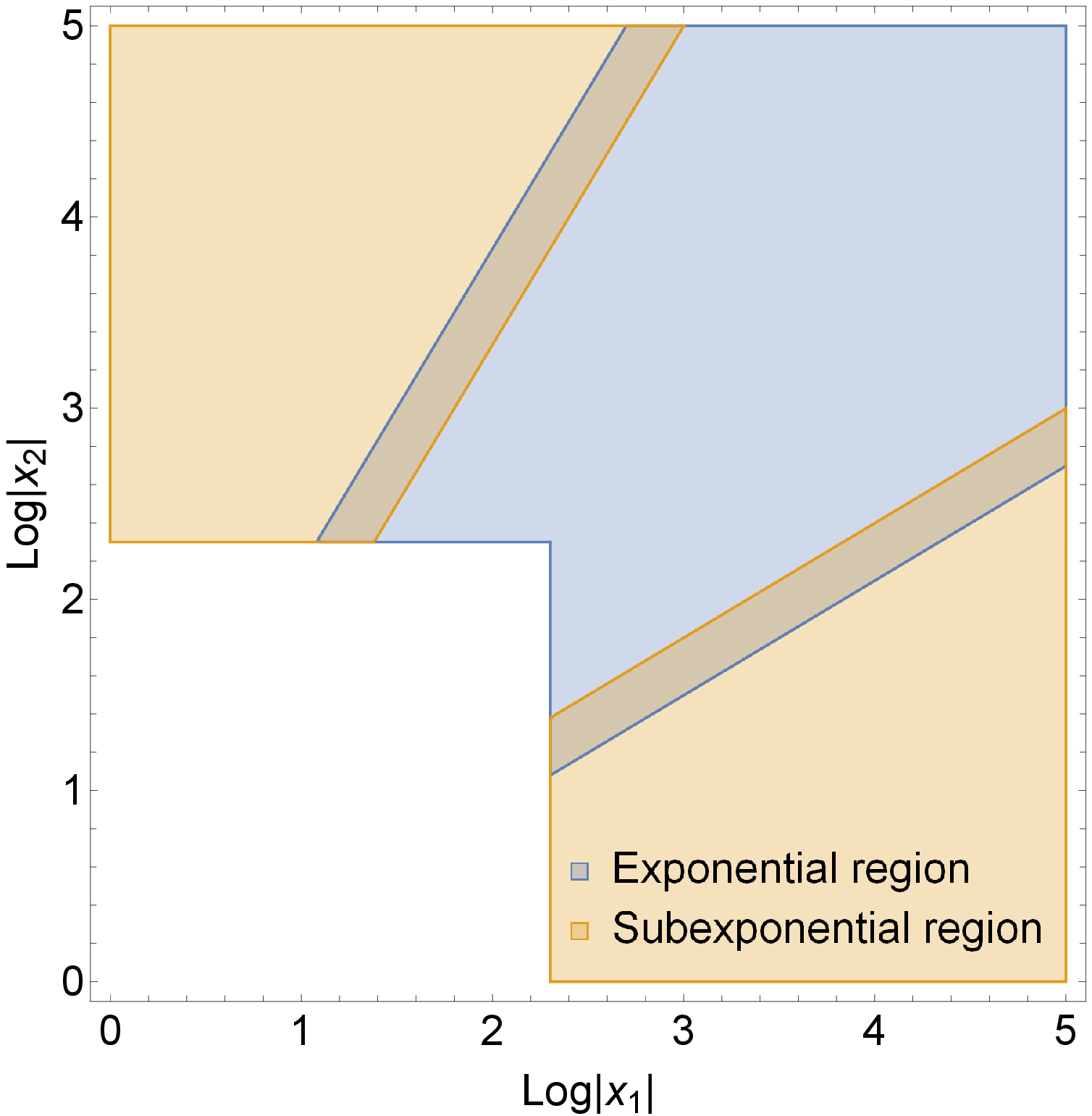}
\caption{Behavior of the ground state eigenfunction for Helium atom at the critical coupling for $\max\{|x_1|,|x_2|\}>200$ with the regions $\{(|x_1|,|x_2|) \,:\, \min\{|x_1|,|x_2|\} > 0.5 \max\{|x_1|,|x_2|\}^{0.6}\}$ (blue), $\{(|x_1|,|x_2|) \,:\, \min\{|x_1|,|x_2|\} <  \max\{|x_1|,|x_2|\}^{0.6}\}$ (amber).}
\centering
\label{fig:Helium_sub_log}
\end{minipage}
\end{figure*}

The difference is not so pronounced in the small region depicted however re-plotting the same regions in logarithmic scale in Figures~\ref{fig:Helium_lin_log}~and~\ref{fig:Helium_sub_log} we see the dramatic difference for large values of $|x|_\infty$.

\end{document}